\documentclass[12pt,a4paper]{article}

\usepackage{amssymb}
\usepackage{hyperref}
\usepackage{parskip}
\usepackage{svg}
\usepackage{authblk}
\usepackage{orcidlink}

\hypersetup{
  colorlinks   = true, 
  urlcolor     = blue, 
  linkcolor    = blue, 
  citecolor   = red 
}

\begin{document}
\renewcommand{\labelenumii}{\arabic{enumi}.\arabic{enumii}}

\title{Discovering and exploring cases of educational source code plagiarism with Dolos}

\author{Rien Maertens\thanks{Corresponding author.} \orcidlink{0000-0002-2927-3032}}
\author{Maarten Van~Neyghem~\orcidlink{0009-0007-1804-4211}}
\author{Maxiem~Geldhof~\orcidlink{0009-0002-6988-7220 }}
\author{Charlotte Van~Petegem~\orcidlink{0000-0003-0779-4897}}
\author{Niko Strijbol~\orcidlink{0000-0002-3161-174X}}
\author{Peter Dawyndt~\orcidlink{0000-0002-1623-90}}
\author{Bart Mesuere~\orcidlink{0000-0003-0610-3441}}

\date{}

\affil{Department of Applied Mathematics Computer Science and Statistics, Ghent University, Ghent, Belgium}

\maketitle

\begin{abstract}
Source code plagiarism is a significant issue in educational practice, and educators need user-friendly tools to cope with such academic dishonesty. This article introduces the latest version of Dolos, a state-of-the-art ecosystem of tools for detecting and preventing plagiarism in educational source code. In this new version, the primary focus has been on enhancing the user experience. Educators can now run the entire plagiarism detection pipeline from a new web app in their browser, eliminating the need for any installation or configuration. Completely redesigned analytics dashboards provide an instant assessment of whether a collection of source files contains suspected cases of plagiarism and how widespread plagiarism is within the collection. The dashboards support hierarchically structured navigation to facilitate zooming in and out of suspect cases. Clusters are an essential new component of the dashboard design, reflecting the observation that plagiarism can occur among larger groups of students. To meet various user needs, the Dolos software stack for source code plagiarism detections now includes a web interface, a JSON application programming interface (API), a command line interface (CLI), a JavaScript library and a preconfigured Docker container. Clear documentation and a free-to-use instance of the web app can be found at \url{https://dolos.ugent.be}. The source code is also available on GitHub.
\end{abstract}

\section{Motivation and significance}

The rise in computer science enrolments \cite{sax2017} and the inclusion of computational thinking and software development in secondary and higher education curricula \cite{balanskatanja2014,uk2013} has resulted in an increase in source code production for classroom assignments.
This worldwide trend comes with its own set of challenges, including source code plagiarism \cite{albluwi2019, pierce2017}. Novak et. al. \cite{novak2019} define source code plagiarism as \textit{“the act of reusing code authored by someone else \textup{[\,\dots]} and failing to adequately acknowledge the fact that the particular source code is not their own”}.
The temptation for students to circumvent learning and to cheat on assessments increases with higher stakes and access to online sources, peer-to-peer communication and generative AI \cite{mccabe1999,ngo2016,ruiperez-valiente2016}.

The migration from paper-based to digital computer science education has increased the use of software tools for detecting source code plagiarism.
These tools aid educators in detecting, proving, and preventing such forms of educational dishonesty by automating the process of finding, comparing, and visualising similar code fragments among large collections of source files.
However, most studies on source code plagiarism rely on unpublished tools that are not or no longer publicly available \cite{novak2019}. Additionally, the process of downloading, installing and running plagiarism detection tools can become tedious and error-prone, which negatively impacts the user experience. Proper interpretation of the results is also a bottleneck \cite{weber-wulff2019}.
This might explain why many educators still refrain from using source code plagiarism detection \cite{chuda2012,culwin2001}. MOSS \cite{schleimer2003} and JPlag \cite{prechelt2002} are currently the most popular free tools for plagiarism detection in educational source code.
However, both tools require local software installations to perform the detection.

For natural language processing, the significance of plagiarism detection is emphasised by the abundance of commercial and free web apps available \cite{chandere2021, jiffriya2021}.
Some of these tools are specifically designed for educational purposes.
They enable educators to conduct plagiarism detection checks directly from their browser, without the need for complex installation procedures or multiple tools.
For source code, various commercial web apps for plagiarism detection exist, such as Codequiry\footnote{\url{https://codequiry.com/}}, Copyleaks\footnote{\url{https://copyleaks.com/code-plagiarism-checker}} and Gradescope\footnote{\url{https://www.gradescope.com/}}. However, none of these apps are fully open-source and free to use.

To fill this gap, we expanded on the initial prototype of Dolos \cite{maertens2022}.
This initial version was already competitive with state-of-the-art tools in terms of performance and prediction accuracy while using a language-agnostic pipeline \cite{maertens2022}.
However, it also needed local installation and its user interface was quite basic.
The latest major release of Dolos (version 2.x) addresses these issues and offers numerous improvements.
The web interface has been redesigned to include new powerful dashboards that allow educators to zoom in from the entire collection, over clusters and pairs, to individual source files.
All visualisations have been significantly improved for better responsiveness and the plagiarism detection pipeline has been optimised for faster runtimes and reduced memory footprints. 
Furthermore, the user experience has also been improved with faster load times, support for anonymisation, automatic programming language detection, highlighting differences between two source files, sharing online dashboards safely with colleagues, and a new packaging strategy for programming language support. 
Finally, a new web app has been developed that obviates the need for local installation. A free-to-use instance of the app is hosted at \url{https://dolos.ugent.be}.
Other instances can be self-hosted to comply with local privacy policies or to use its API for seamless integration into online learning environments.
While the command line interface (CLI) from the first version is still supported, all new and improved features are now also accessible from the new web interface.

\section{Illustrative example}

This section provides instructions on how to use the Dolos web app to detect plagiarism in a collection of programming assignment submissions. From an educator’s perspective, the process involves two steps: \textit{i)} uploading source files and \textit{ii)} checking dashboards for suspected cases.
To follow along, it is recommended to use the free-to-use instance hosted at Ghent University (\url{https://dolos.ugent.be}) with either your own collection of submissions or our sample dataset.
You can also take a guided video tour at \url{https://dolos.ugent.be/tour}.

For more information on how to install and run the CLI locally, self-host a local instance of the web app using Docker, or use the JavaScript library directly, please refer to the online documentation at \url{https://dolos.ugent.be/docs}.
Additionally, a guide to adding new programming languages is available.

\subsection*{Step 1: data submission}

\begin{figure}
\includegraphics[width=\columnwidth]{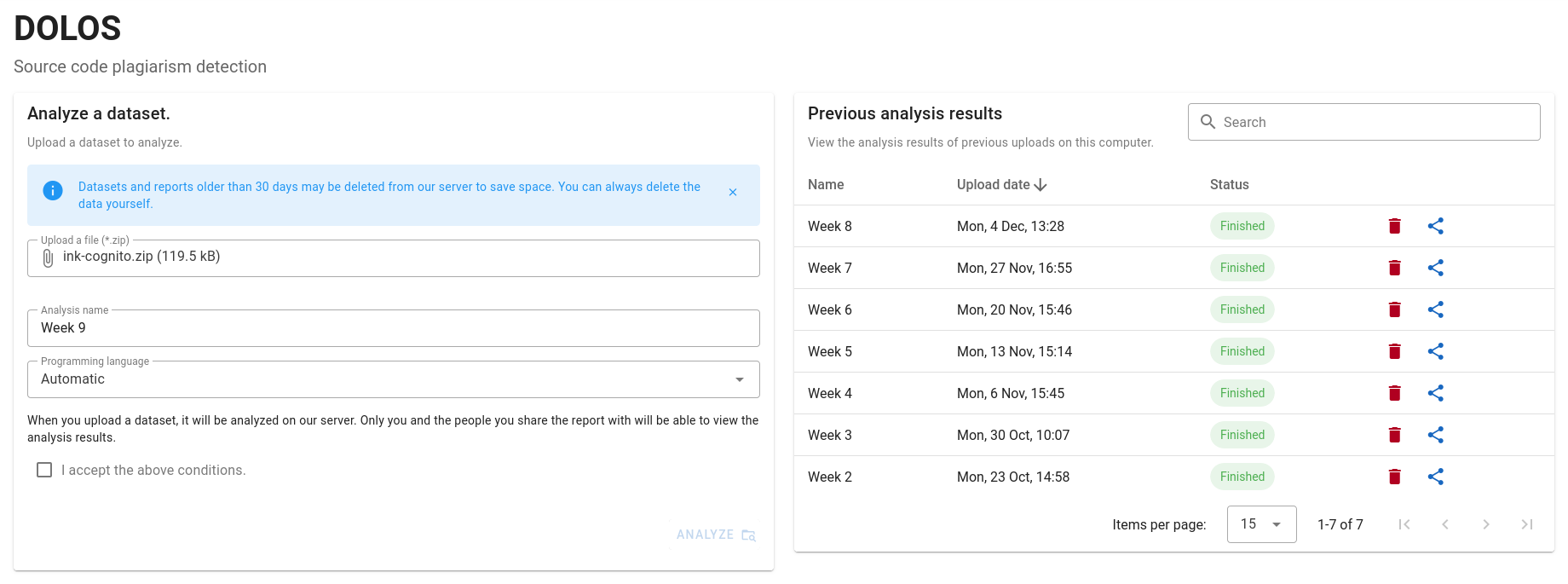}
\caption{
Launchpad of the Dolos web app. Left panel: upload form for submitting a new collection of source files. Right panel: searchable table for accessing, deleting and sharing previously submitted collections.
}
\label{fig:launchpad}
\end{figure}

The web app’s launchpad consists of two panels (Figure \ref{fig:launchpad}). The left panel features an upload form for submitting new collections of source files, while the right panel contains a searchable table for retrieving previously submitted collections and reviewing their analysis results.

To submit a new collection, begin by selecting a ZIP archive containing the source files from the local file system.
The archive may also contain a CSV-formatted file with metadata such as submission timestamps, authors and free-form labels.
Dolos currently supports the simple CSV-format exported by Dodona \cite{vanpetegem2023}.
It is important to note that all source files must share the same programming language.
The app will automatically detect the language during file selection, but it can be manually overridden by selecting from a drop-down list of all supported programming languages.
The app will also suggest a name for the collection, which can be edited as needed.

Upon submission, the app launches a server-side job that executes the source code similarity analysis pipeline (Section \ref{sec:algorithms}).
Jobs usually finish within a few seconds, and the results are then accessible for further examination.
The user interface has been intentionally designed to be minimalist by running the analysis pipeline with finely tuned parameters that cannot be altered.
Advanced users can run Dolos from its CLI or JavaScript library. 
For collections containing over than 1000 files, files with over 1000 lines of code, or when integrating Dolos into an automated pipeline, we recommend this approach.

Each uploaded collection of source and metadata files is stored server-side, along with a submission timestamp and associated analysis results. 
The app does not rely on user accounts to manage collections. 
Instead, each analysed collection is assigned a unique \textbf{secret key} that is stored in the browser’s local storage.
We take special care to prevent malicious access by ensuring that browsers do not leak secret keys when displaying URLs.
These keys are all that the app needs to retrieve server-side metadata from previously analysed collections in a searchable table.
This table allows for easy access to analysis results, which can be shared with colleagues or deleted both client and server-side. The instance hosted at Ghent University guarantees a 30-day retention period.
After this period, analysed collections are periodically deleted from the server.
Lost keys (e.g. when deleting browser data) or scheduled server-side deletion are not too painful.
The results can be easily and quickly reproduced by re-running the analysis from a new submission.

\subsection*{Step 2: exploring analysis results}

Like other plagiarism detection tools, Dolos’ server-side analysis pipeline merely automates the detection of highly similar code fragments shared between source files and calculates pairwise similarities between each pair of files in the collection.
However, gathering enough convincing evidence is undoubtedly the most challenging aspect of dealing with educational source code plagiarism.
Students are typically not caught in the act. This task is challenging to fully automate, but the web app assists the educator’s expert eye with new and carefully crafted \textbf{plagiarism analytics dashboards}.

Dashboards are provided for various subsets of source files in the \textbf{collection}: the complete collection, a \textbf{cluster} of files, a \textbf{pair} of files, and a \textbf{single} file. 
This creates a hierarchical structure of linked dashboards at different zoom levels. 
Each dashboard offers custom analytics and visualisations to compare files within the subset and with files outside the subset.
Moving between linked dashboards provides a natural zooming experience when investigating suspected cases of plagiarism.
The left-hand navigation bar also includes searchable tables for subsets at each zoom level (clusters, pairs and individual files).
These tables allow for comparison of subsets and enable drilling down into specific subsets.

\begin{figure}
\includegraphics[width=\columnwidth]{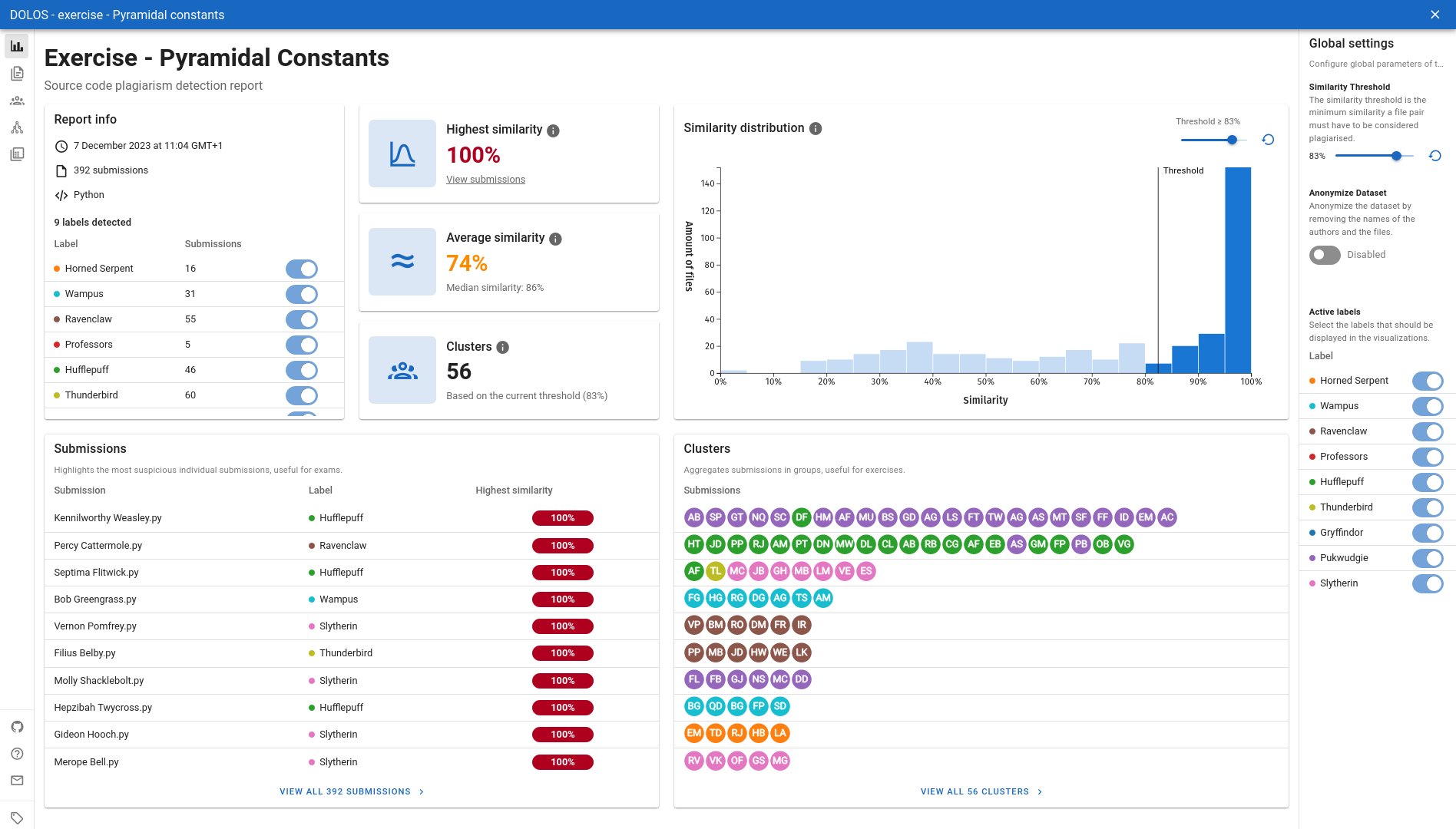}
\caption{The overview dashboard’s analytics and visualisations summarise the plagiarism detection results. This specific report suggests that plagiarism is prevalent in this publicly available collection of source files. The collection info card (top left) displays basic statistics about the collection being analysed. Colour codes for the highest and average pairwise similarities (top centre and bottom left) between files indicate the level of suspicion of plagiarism, ranging from low (green), to average (orange) and high (red). The histogram (top right) and a list (bottom left) display the global similarity with the nearest neighbour of each source file. The composition of clusters (bottom right) represents the source files as circles marked with an acronym derived from their author name, and coloured according to their label. Student subjects are used as labels for this collection of source files. The individual files (bottom left) and clusters (bottom right) are ranked by decreasing suspicion of plagiarism. The web app uses a simple heuristic to determine an appropriate initial similarity threshold for clustering. This threshold can be modified either in the histogram (top right panel) or in the global settings (activated on the far right of the top navigation bar). All dashboards also have a shared setting that anonymises analytics and visualisations (useful for in-class demonstrations) and a label-based filtering for the collection of source files.}
\label{fig:overview}
\end{figure}

The exploration of the complete collection starts at an \textbf{overview dashboard} (Figure \ref{fig:overview}).
Its analytics and visualisations provide an immediate impression of whether the collection contains suspected cases of plagiarism and the extent of plagiarism within the collection. 
Clues can be found, for example, by contrasting the highest and average pairwise similarities between files, relating source files to their nearest neighbour in terms of global similarity (both available as a histogram and a list), and inspecting the number and size of file clusters.

\begin{figure}
\includegraphics[width=\columnwidth]{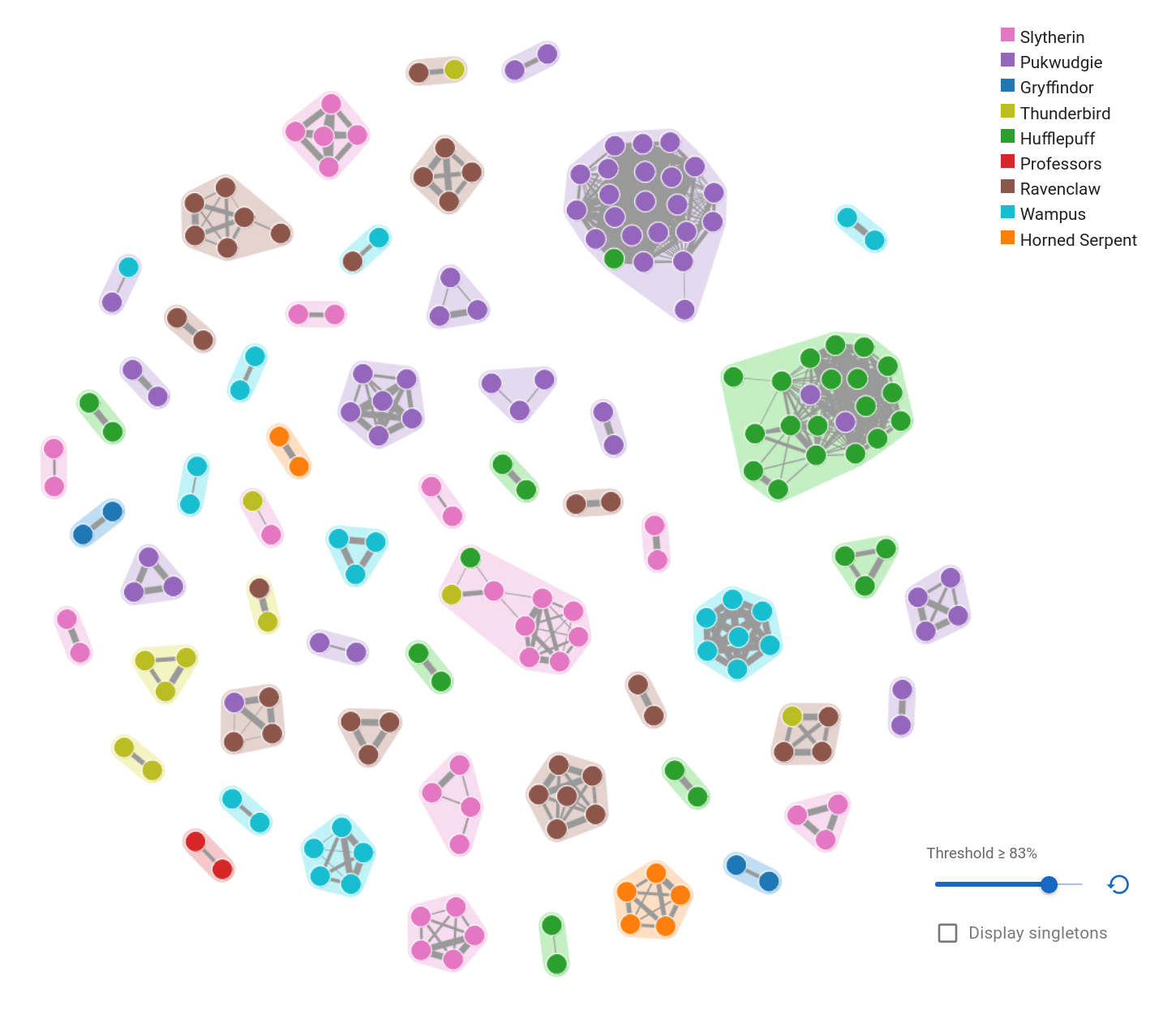}
\caption{
Graph showing suspected cases of plagiarism within the same collection of source files used for Figure \ref{fig:overview}. Each node represents a source file and has a colour that corresponds to its file labels. The legend (top right) can be used to include or exclude files from the graph by label. Edges connect nodes whose pairwise similarity exceeds an adjustable threshold (bottom right), set at 83\% global similarity. Clusters of connected nodes are grouped within regions whose background colour reflects the dominant colour of the cluster nodes. Source files are excluded from the graph view if their global similarity with the nearest neighbour falls below the threshold (i.e. nodes not connected by an edge to any other node in the graph), unless the ``Display singletons" option (bottom right) is enabled.
}
\label{fig:graph}
\end{figure}

The same underlying goal led us to visualise the hierarchically structured subsets of the collection as a \textbf{plagiarism graph} (Figure \ref{fig:graph}).
The initial view shows only suspect files (as nodes coloured by label), pairs (as edges drawn between nodes whose global similarity exceeds a similarity threshold), and clusters (coloured regions that group nodes connected by edges). 
There’s also an option to include all files in the collection in the graph display. 
This gives a better understanding of the sparsity of the solution space for a programming exercise and the prevalence of plagiarism within the submitted solutions.
The graph view is provided on a separate page due to space limitations, where it would logically fit on the overview dashboard.

Existing source code plagiarism tools traditionally only report potential plagiarism from the perspective of individual files or file pairs.
However, larger groups of collaborating students quickly result in an unmanageable list of file pairs (e.g. 10 students result in 45 file pairs), which may be scattered across a list of reported file pairs.
However, seeing the same data visualised as a clustered graph feels very intuitive. 
As a result, the \textbf{cluster} concept is now an integral part of the Dolos dashboard design as a separate hierarchical level.
This feature helps distinguish between peer-to-peer plagiarism events (two students sharing code) and broadcast events (larger groups of students sharing code, e.g. via social media). 
The cluster dashboard reconstructs the distribution timeline based on submission timestamps. 
This feature is useful for tracking the original author or observing how the distribution process has evolved over time.

\begin{figure}
\includegraphics[width=\columnwidth]{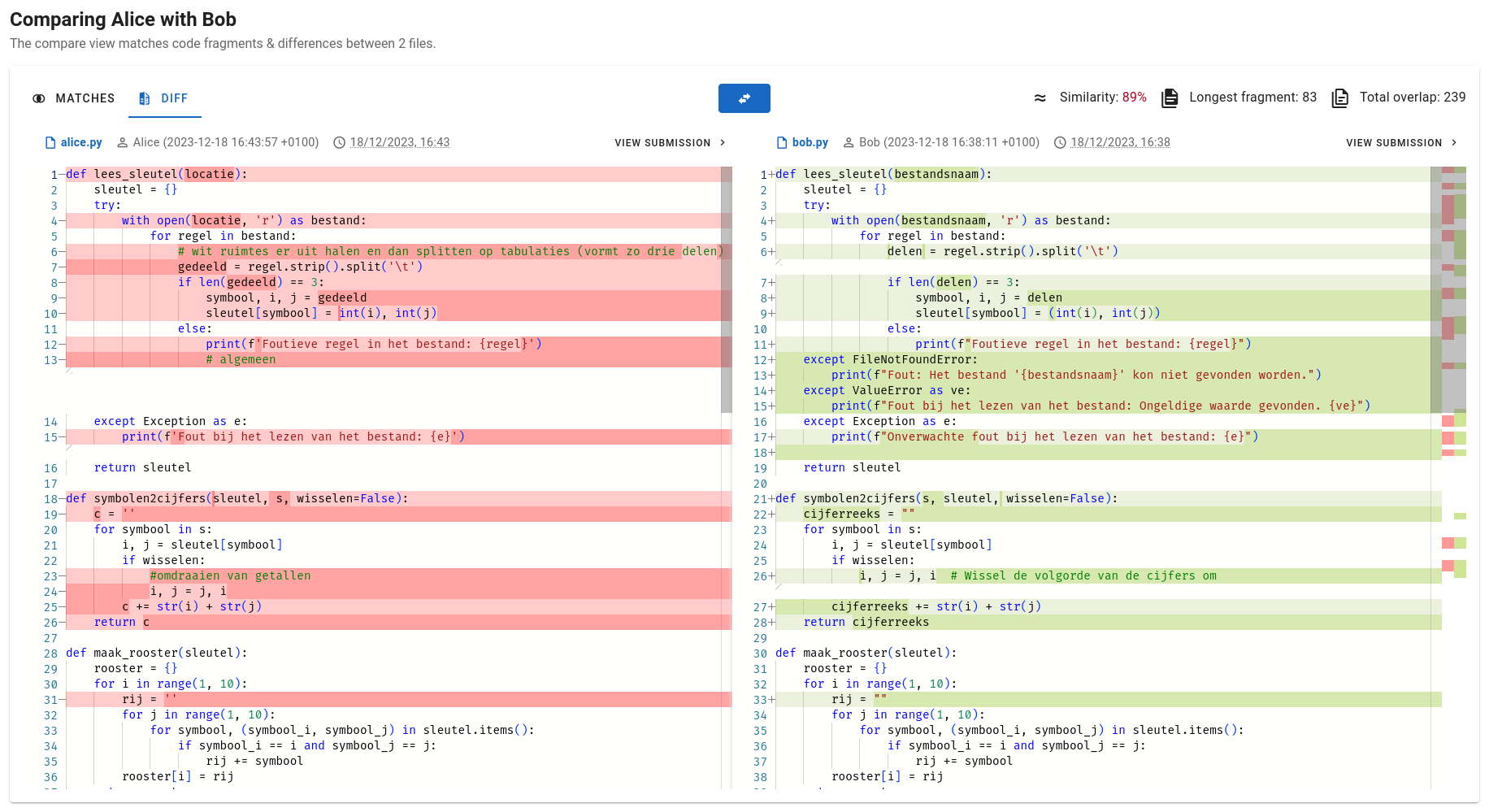}
\caption{
The new diff view highlights the differences in the dashboard for comparing two files. In this particular case, the two solutions are almost identical, with only minor syntactic differences such as parameter and variable names, comments and string quotes. It is possible that one of the students made these changes in an attempt to disguise plagiarism.
}
\label{fig:compare}
\end{figure}

The \textbf{pair dashboard} displays two source files side by side (Figure \ref{fig:compare}).
This feature enables educators to differentiate between genuine and false (or questionable) cases of plagiarism.
It also assists them in identifying adequate and conclusive evidence that high global similarity or lengthy shared fragments are not coincidental. 
It is worth noting that many students intentionally employ various obfuscation techniques \cite{novak2019} to conceal that they have copied someone else’s code.
Both plagiarism detection pipelines and educators must try to see through this.
The pair dashboard offers two views: one highlights matching fragments found by the Dolos plagiarism detection pipeline (Section \ref{sec:algorithms}), while the other highlights differences found by string alignment \cite{myers1986}.
Matching fragments provide more insight when source files are globally less similar, or when blocks of code have been rearranged.
The diff view helps to highlight small syntactic changes when source files are very similar.
The Dolos pipeline masks some of these changes to see through known obfuscation patterns. When educators land on a pair dashboard, the app automatically selects the most relevant view for the two source files at hand.

All dashboards share three global settings, which can be modified in a dedicated panel (expanded from the far right of the top navigation bar; Figure \ref{fig:overview}) or in some panels within the dashboards themselves.
Suspect pairs and clusters are delineated based on a global \textbf{similarity threshold}. Dolos employs a simple heuristic to automatically determine an appropriate initial value for this threshold (Section \ref{sec:algorithms}).
All analytics and visualisations can be \textbf{anonymised} to present dashboards in a privacy-friendly manner.
Discussing the impact and consequences of plagiarism with students could be part of a preventive strategy \cite{berrezueta-guzman2023, maertens2022}.
\textbf{Label-based filtering} is used to control which subset of the total collection is considered by the dashboards.

\section{Software description}

All Dolos source code is available in a public monorepo on GitHub\footnote{\url{https://github.com/dodona-edu/dolos}} and in the Zenodo software repository \cite{maertens2023}.
This section describes the software architecture of the web app (version 2.x) and explains some of the algorithmic design decisions. It is intended for researchers, developers, and power users who wish to understand the system internals, reuse components in isolation or contribute to the project.

\subsection{Software architecture}
\begin{figure}
\includegraphics[width=\columnwidth]{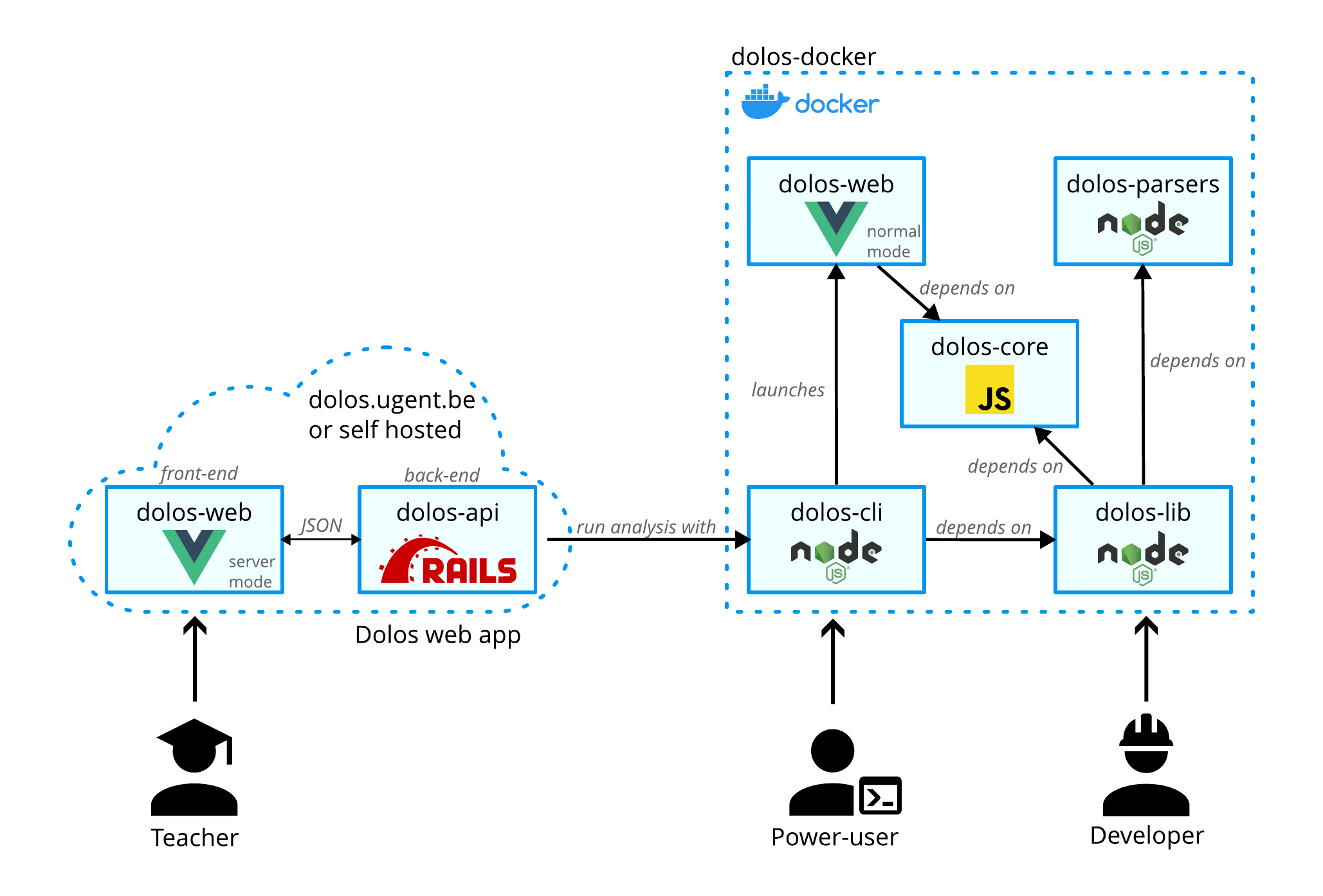}
\caption{
Diagram of the different components in the Dolos ecosystem and their relationships (Dolos version 2.x). Some components can be used in isolation, as shown by the three users interacting with the components. External dependencies and standalone documentation pages (\texttt{dolos-docs}) have been excluded.
}
\label{fig:ecosystem}
\end{figure}

These software components make up the Dolos web app (Figure \ref{fig:ecosystem}):

\begin{itemize}
\item[\texttt{dolos-core}] implements core algorithms of the source code similarity analysis pipeline.
An original TypeScript implementation of the winnowing algorithm \cite{schleimer2003} that is transpiled into a pure ECMAScript Module (ESM) without external dependencies.
The ESM package can be executed on any platform that provides a JavaScript runtime engine (web browser, Node.js).

\item[\texttt{dolos-parsers}] collection of tree-sitter parsers \cite{brunsfeld2024} for major programming languages bundled in a single package.
Aggregates a collection of Git submodules with a \texttt{node-gyp} configuration to these build parsers and create a single JavaScript package with the resulting node bindings.
A custom module aggregating all parsers allows faster integration of additional programming languages into Dolos and keeps supported languages up to date.
Avoids dependencies on maintainers of individual parsers to publish new releases on npm.

\item[\texttt{dolos-lib}] Node.js\footnote{\url{https://nodejs.org/}} library for reading source files, parsing source code (depends on \texttt{dolos-parsers}), and generating plagiarism reports (depends on \texttt{dolos-core}).
Supports integration of plagiarism detection into online learning environments.
Re-exports algorithms implemented in \texttt{dolos-core}.

\item[\texttt{dolos-web}] web interface implemented on top of the Vue 3\footnote{\url{https://v3.vuejs.org}} JavaScript framework. Provides clean and consistent UX/UI by using Vuetify\footnote{\url{https://v3.vuetifyjs.com}} components.
Includes D3-based \cite{bostock2011a} interactive visualisations from dashboards.
Depends on \texttt{dolos-core} for client-side execution of some plagiarism analysis pipeline steps (in browser) to keep the app responsive and interactive.
May be built in \textit{normal mode} to generate dashboards from an external run of the plagiarism analysis pipeline (used by \texttt{dolos-cli}).
May be built in \textit{server mode} to add upload functionality onto a \textit{normal mode} build used in the Dolos web app, interacting with the \texttt{dolos-api}.

\item[\texttt{dolos-cli}] Node.js command line interface (CLI) for plagiarism detection functionalities provided by \texttt{dolos-lib}.
Results from the analysis pipeline can be displayed in the terminal, exported to CSV-files, or launched as dashboards in the browser (depends on \texttt{dolos-web}).

\item[\texttt{dolos-api}] Ruby on Rails\footnote{\url{https://rubyonrails.org/}} web server exposing an application programming interface (API) for plagiarism detection functionalities provided by Dolos. Results are returned in JSON format.
For proper sandboxing, each new request uploads source files and runs \texttt{dolos-cli} in its own Docker container (\texttt{dolos-docker}). The analysis results are stored server-side. The \texttt{dolos-web} web interface built in \textit{server mode} communicates with this API to upload datasets, fetch report information, and display the results.

\end{itemize}

The Dolos documentation website’s (\url{https://dolos.ugent.be}) source code is included in the \texttt{dolos-docs} module.
The \texttt{dolos-docker} module contains a Docker\footnote{\url{https://docker.com}} container pre-installed with the Dolos CLI (\texttt{dolos-cli} component).
For each new release, a new version of the \texttt{dolos-docker} package is automatically published in the GitHub container registry (\url{https://ghcr.io/dodona-edu/dolos}). Additionally, new versions of the \texttt{dolos-core}, \texttt{dolos-parsers}, \texttt{dolos-lib}, \texttt{dolos-web} and \texttt{dolos-cli} packages are automatically published on npm under the \texttt{@dodona} scope: \texttt{@dodona/dolos-core}, \texttt{@dodona/dolos-parsers}, \texttt{@dodona/dolos-lib}, \texttt{@dodona/dolos-web}, with \texttt{@dodona/dolos} providing the \texttt{dolos-cli} package.

\subsection{Algorithms}
\label{sec:algorithms}

In this section we highlight some of our algorithmic design choices.

\subsection*{Programming language support}

The pipeline for plagiarism detection uses concrete syntax trees (CSTs) as its internal representation of source files.
To protect against known syntactic plagiarism obfuscation patterns, tokens such as identifiers, string content, and comments are masked.
The CSTs are then serialised and passed through a programming language agnostic string matching pipeline. This is accomplished by delegating code parsing and CST generation to tree-sitter’s\footnote{\url{ https://tree-sitter.github.io/tree-sitter/}} generalised parser generators \cite{brunsfeld2024}. Currently, tree-sitter parsers are available for more than 130 programming languages, with new languages being added regularly.

In theory, Dolos language support depends solely on the availability of tree-sitter parsers.
However, in practice, the npm JavaScript package registry, which provides the easiest way to install parsers, does not always contain the latest version of a parser, if it contains it at all.
Most parsers are developed and maintained by third-party developers rather than the core tree-sitter team, and each major release of Node.js or the tree-sitter core library requires new npm releases of the parsers to avoid incompatibilities.
The \texttt{dolos-parsers} component solves this problem by using Git submodules to embed source code repositories of individual parsers and bundles them into a single JavaScript module.
This reduces the workload for maintainers of tree-sitter parsers and avoids dependencies on their npm release strategy.
We are actively extending \texttt{dolos-parsers} with new language parsers, and encourage users to suggest new languages on the GitHub repository if they do not find the language of their choice in the list of officially supported languages..
The outcome is a comprehensive collection of parsers for languages officially supported by Dolos.

\subsection*{Tracing shared fragments and computing pairwise similarities}

After converting the source files into CSTs, masking and serialisation, Dolos uses a custom TypeScript implementation of the winnowing algorithm \cite{schleimer2003} to extract fingerprints from the token streams of each source file.
The pairwise global similarity computation is based on substrings of fingerprints shared between the two source files, making it immune to code block reordering.
The web app also visualises the corresponding matching fragments in its pair dashboard.
This aims to assist educators in understanding how students may have rearranged code blocks in an attempt to conceal plagiarism.

The winnowing algorithm begins by hashing each $k$-gram (overlapping substrings of $k$ consecutive tokens) into a fingerprint (an integer hash), taking the token stream extracted from a source file as input. 
To ensure efficiency, a fast rolling hash algorithm has been implemented, taking into account the lack of a specific integer datatype in JavaScript.
Absolute values of fingerprints are kept strictly below $2^53$ to force the internals of the JavaScript engine to rely only on efficient integer computations for all hash operations.
Fingerprints are sampled at a rate of approximately one per window of w fingerprints to reduce memory usage.
Retaining the fingerprint with the lowest hash value from each window consistently ensures that common code fragments are highly likely to be preserved as common substrings of fingerprints.
For further details, please refer to the papers on the original winnowing algorithm \cite{schleimer2003} and the first Dolos version \cite{maertens2022}.

Calculating cross-comparison metrics (shared fragments, total overlap, similarity) for each pair of source files in a collection is the most performance-critical aspect of the plagiarism detection pipeline.
In the latest version of Dolos, we have replaced the previous method of identifying shared fragments between a pair of source files with a faster algorithm that finds all longest common substrings. 
This has significantly reduced the computation time and memory usage of the entire pipeline.
However, Dolos’s processing speed is still dependent on the number of files and the total file length, resulting in a quadratic and linear relationship, respectively. As a result, when processing collections of more than 1000 source files or files with more than 1000 lines of code, Dolos becomes noticeably slower.
To alleviate this issue, the two parameters0 $k$ and $w$ of the winnowing algorithm can be relaxed.
However, to avoid complicating the user experience, we have decided not to support this option in the web app.
Instead, we are currently investigating the possibility of achieving near-linear time complexity by using generalised suffix trees and more memory-friendly variations.
This approach aims to improve the efficiency of the process.

\subsection*{Automatic threshold detection}

A global similarity threshold determines which pairs of source files are suspect (i.e. connected by an edge in the plagiarism graph) and how files are clustered in the web app dashboards.
The goal is to avoid false positives (suspected or clustered files that are not plagiarised) and false negatives (plagiarism events that go unnoticed).
However, selecting the optimal threshold can be challenging.
It is highly dependent on various factors, including the expected file size, the number of students, the programming language used, and the diversity of the solution space.
The threshold is therefore highly dependent on the collection of source files being analysed.
To assist educators, the web app automatically infers an initial threshold when dashboards are launched for a given collection.
This is done under the assumption that all the files are solutions to the same programming assignment. 
If any of these solutions have been plagiarised, the histogram of global similarities between the files and their nearest neighbours is expected to show the superposition of two Gaussian distributions.
One is centred in the lower half of the similarity interval and corresponds to non-plagiarised files. 
The other is centred near the maximum of the similarity interval and corresponds to plagiarised files.
Dolos uses a heuristic to estimate the point where the two distributions intersect as the initial threshold. 

\section{Impact}

In May 2023, following a “release often/release early” strategy, Ghent University (Belgium) started hosting a first standalone instance of the Dolos web app.
At the time of writing (February 2024) this preview version alone has scanned over 2700 collections of source files for possible cases of plagiarism.
The significance of source code plagiarism in education is further highlighted by the fact that Dolos has received over 180 stars on GitHub from people from around the world.
The code repository also had 49 issues or discussions opened by users outside the core development team to report bugs, ask questions or suggest features for unsupported use cases.

Industry players have begun integrating the Dolos web app into their online learning environments.
Codio\footnote{\url{https://codio.com}}, an online platform that supports computer science courses, recently switched from using MOSS and JPlag for source code plagiarism detection to a self-hosted instance of Dolos. 
They justify this decision on their website, stating that: \textit{“Plagiarism detection systems available such as MOSS and JPlag were not developed for university programming courses. 
Therefore, they can require considerable effort to submit large files of student code projects and to interpret the results. 
Codio integrates the Dolos plagiarism detection system developed by CS educators for programming courses. 
This integration provides instructors with enough data and analysis for a lecturer to make a conclusive, final decision. 
The burden of project data preparation and submission to remote systems such as MOSS and JPlag is removed.
The result is a single-click process for the lecturer or teacher.”}

Software components of the Dolos code similarity and clustering pipeline are also being used beyond the original application domain of educational source code plagiarism detection. 
For instance, a study on the prevalence of large language models (LLMs) violating software copyright, Yu et. al. \cite{yu2023} used Dolos to compare original copyrighted source code with LLM-generated code for. Dolos has also been used for malware detection (personal communication), where $k$-gram analysis is commonly used to classify computer viruses \cite{gandotra2014}.

\section{Conclusions}

The latest major release of Dolos (version 2.x) includes a free and open-source web app for educators to detect plagiarism in educational source code.
This novel app can be run directly from the browser without any installation or configuration.
It is built on top of a state-of-the-art source code similarity detection pipeline that has been optimised for speed and memory consumption. 
The app supports numerous programming languages out of the box, and the procedure for adding new language parsers has been enhanced.
It offers a well-designed set of dashboards for plagiarism analytics. 
The hierarchical structure of the dashboards enables a thorough examination of suspected plagiarism cases within a collection of source files. 
Identifying clusters of source files helps comprehend the distribution of plagiarism incidents among groups of students.
A comparison of source files side by side can help to identify conclusive evidence that high code similarity is not a coincidence.

Dolos primarily focuses on detecting source code plagiarism in educational settings.
However, it has also been utilised for other code similarity and clustering applications, such as malware analysis and generative AI research.
We offer comprehensive documentation for power users who wish to host an instance of the web app, integrate plagiarism detection into external learning platforms using its JSON API or JavaScript library, or perform source code similarity analysis from the command line.
Dolos’ roadmap includes further research into the use of advanced index structures to enable fast scanning of more and longer source files. 
Additionally, we want to provide specific support for multi-file student projects and take into account the additional longitudinal dimension of students submitting multiple solutions to the same programming exercise. 
Collaboration on these issues is welcome, and we would be happy to hear about other use cases.

\section*{Acknowledgements}

Dolos is part of the ecosystem surrounding the Dodona online learning platform\footnote{\url{https://dodona.ugent.be}} \cite{vanpetegem2023}.
Team Dodona expresses gratitude for the financial support provided by Ghent University (UGent, Belgium) and the Flemish Government (Belgium, Voorsprongfonds) through various grants for innovation in education.
Additionally, we thank UGent for hosting a free-to-use instance of the Dolos web app (\url{https://dolos.ugent.be}).
This work was partially supported by the Research Foundation — Flanders (FWO) for ELIXIR Belgium (I002819N). 

We thank UGent for awarding us with the 2018 Minerva Award for our contribution to active learning and innovation in education through the development of Dodona.
We are also proud to have received the 2022 Flanders Digital Award from the Flemish Government (Belgium) for providing high-quality education to every student through Dodona.
We appreciate all users who reported issues, shared use cases and provided constructive feedback.

\bibliographystyle{elsarticle-num} 
\bibliography{main}

\end{document}